# Femtosecond spectroscopy with vacuum ultraviolet pulse pairs

T. K. Allison,[1,2,*] T. W. Wright,[2,3] A. M. Stooke,[1] C. Khurmi,[2] J. van Tilborg,[2] Y. Liu,[1,2] R. W. Falcone,[1,2] and A. Belkacem[2]

[1]Department of Physics, University of California at Berkeley, Berkeley, California 94720, USA
[2]Ultrafast X-ray Science Laboratory, Lawrence Berkeley National Laboratory, Berkeley, California, 94720, USA
[3]Department of Chemistry, University of California at Davis, Davis, California, 95616, USA
*Corresponding author: tka3@jila.colorado.edu

We combine different wavelengths from an intense high-order harmonics source with variable delay at the focus of a split-mirror interferometer to conduct pump–probe experiments on gas-phase molecules. We report measurements of the time resolution (<44 fs) and spatial profiles (4 $\mu$m × 12 $\mu$m) at the focus of the apparatus. We demonstrate the utility of this two-color, high-order-harmonic technique by time resolving molecular hydrogen elimination from $C_2H_4$ excited into its absorption band at 161 nm.   © 2010 Optical Society of America
  OCIS codes:   300.6540, 320.2250.

Pump–probe spectroscopies with femtosecond pulses have become standard and powerful tools for following the ultrafast dynamics of atoms, molecules, and condensed matter in real time. Most of these studies have utilized pulses in the visible and IR regions of the spectrum where femtosecond laser technology is mature. Fundamentally, atoms and molecules have the bulk of their oscillator strength in the vacuum ultraviolet (VUV) and extreme ultraviolet (XUV), between the wavelengths of 200 and 30 nm, but both producing femtosecond pulses of sufficient intensity and combining them with variable delay in this spectral range poses technical challenges. High-order-harmonic generation (HHG) [1] can provide ultrashort pulses in the VUV and XUV, and many groups are now using HHG combined with multiphoton excitation from long-wavelength (400–800 nm) pulses for time-resolved photoionization studies [2–4]. However, to study molecular dynamics without the complications of multiphoton or strong field physics, one would like both pump and probe pulses to act as weak perturbations and interact with a system through the absorption of only one photon each. In this Letter, we describe and characterize a tabletop apparatus for conducting femtosecond pump–probe spectroscopy with VUV and XUV pulses and report time-resolved measurements of the molecular hydrogen elimination from $C_2H_4$ excited at 161 nm.

The apparatus is depicted in Fig. 1. High-order harmonics of 807 nm are generated with a repetition rate of 10 Hz by loosely focusing ($f = 6$ m) 30 mJ, 50 fs laser pulses into a 5 cm gas cell with laser drilled pinholes. The cell is filled with 2.4 Torr of Xe and scanned through the focus to optimize the harmonic yield. Greater than $10^{10}$ photons/harmonic/shot emerge from the gas cell in orders 11–15 ($h\nu = 17$–23 eV) along with more than 200 nJ in the fifth harmonic ($h\nu = 7.7$ eV); for more details, see [5]. The harmonic and fundamental beams are allowed to diverge for 3 m, where they are incident on a superpolished silicon mirror set at the 800 nm Brewster angle (75°). The silicon mirror removes the fundamental and reflects the harmonics [6].

Pump/probe delay is achieved with a split-mirror interferometer (SMI) similar to that described in [7]. The harmonics are focused into a pulsed molecular beam by two D-shaped spherical concave mirrors ($r = 20$ cm) at normal incidence. Photoions from the focal region are selected with a 500 $\mu$m aperture and measured with a time-of-flight (TOF) ion mass spectrometer. The probe arm mirror is mounted on a piezoelectric translation stage to produce a time delay. The pump arm mirror is mounted on a tip/tilt flexure with motorized actuators to optimize the spatial overlap of the two foci. The pitch and yaw angles of the pump arm mirror are recorded by hermetically sealed linear variable differential transducers mounted to the flexure assembly. Wavelength selection in each arm of the SMI is achieved through a combination of transmission filters and coatings on the two D-shaped mirrors.

We chose to use the fifth harmonic at 161 nm as the pump wavelength. We characterized the spatial profile of the fifth harmonic at focus by selecting the fifth harmonic in both arms of the SMI and scanning the pitch and yaw angles of the pump arm mirror while monitoring the two photon ionization of $C_2H_4$ target gas. A spectral bandpass filter (Acton Research 160 N on 0.2 mm substrate) was inserted in both arms of the SMI to isolate the fifth harmonic, and the delay stage was set so that the pulses from both arms arrived at focus simultaneously. The spatial autocorrelations in the vertical and horizontal directions are shown in Fig. 2 and fit with Gaussians of 6 and 18 µm FWHM, respectively. The main source of the worse beam quality in the horizontal plane is likely the sharp edge in each half-beam produced by the gap between the mirrors [8]. Another source could be the larger illuminated length of the Si mirror, with an imperfect surface figure, in the horizontal plane. The incident beam size was measured to be 3 mm, which would give a 2 µm diffraction-limited spot size. Assuming the vertical cross correlation represents the cross correlation of two Gaussian beams, the vertical beam size at focus is $6~\mu m/\sqrt{2} = 4.2~\mu m$, corresponding to an $M_y^2$ for the fifth harmonic of roughly 2.

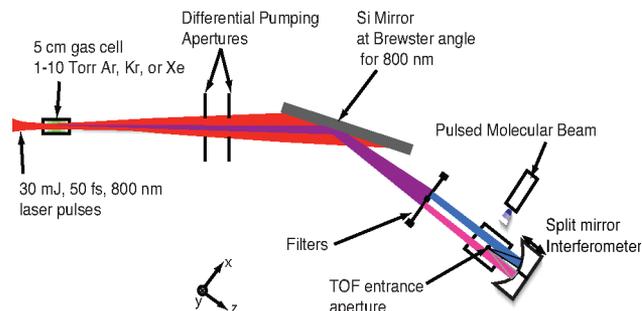

Fig. 1.  VUV/XUV pump–probe apparatus: high-order harmonics are recombined with a variable delay at the common focus of two spherical concave mirrors. Photoions are extracted from the focal region through a 500 $\mu$m pinhole with a strong electric field in the $y$ direction.

Time-resolved measurements of photodissociation reactions of the form $h\nu + AB \rightarrow A + B$ are conducted by probing with XUV harmonics with photon energies above the ionization potentials of the products and recording the yield of $B^+$. In the probe arm, we inserted a thin tin (Sn) foil to select harmonics 11–15. The probe arm mirror consisted of a magnetron sputtered 30 nm $B_4C$ coating ($R \approx 30\%$ for $h\nu = 17$–$25$ eV) on a fused silica substrate. The pump arm mirror was a multilayer dielectric stack (Layertech Gmbh) designed to reflect and focus the fifth harmonic ($R > 90\%$) and transmit the third harmonic and fundamental. We also inserted a 230-µm-thick $CaF_2$ window in the pump arm, which serves three purposes: (i) it absorbs all harmonic orders above 5, (ii) the group-velocity dispersion (GVD) of the $CaF_2$ window provides dispersion compensation for the negative chirp of fifth-harmonic pulse that emerges from the gas cell due to the atomic dipole phase intrinsic to the HHG process ([9] and references therein), and (iii) the GVD of the window temporally separates the desired fifth harmonic from unwanted remnants of the third harmonic and fundamental that reflect off the dielectric mirror. Thus, residual third harmonic and fundamental light pulses arrive at the focus more than 300 fs earlier than the fifth harmonic and being nonresonant with our target gas, pass through unabsorbed and do not corrupt the experiment.

The mirrors and filters can be moved perpendicular to the beam path to adjust the power in the pump and probe arms. Large excitation fractions were possible with the pump pulse. For example, for $C_2H_4$ target gas with a photoabsorption cross section of 25 Mb at 161 nm, we observed the ability to deplete the observed $C_2H_4^+$ signal [as in Fig. 3(b)] by more than 10%, while the XUV probe produced a signal of $\sim 100$ $C_2H_4^+$ ions per shot.

The time resolution for the VUV pump/XUV probe configuration is estimated from the photodissociation of water vapor. Upon excitation to the $\tilde{A}^1B_1$ state with 160 nm light, the $H_2O$ molecule rapidly dissociates to $OH + H$ along a purely repulsive potential [10]. The step-like increase of the $OH^+$ signal as the molecule dissociates is shown in Fig. 3(a). We assume that this very fast reaction occurs within our time resolution and fit it with an error function to give an upper limit for the finite instrument response (FIR) of $44^{+15}_{-10}$ fs FWHM (convolution of pump and probe).

The photodissociation of ethylene ($C_2H_4$) excited at 161 nm represents a more interesting case. The first absorption band of ethylene, peaking at around 165 nm, is dominated by the $\pi \rightarrow \pi^*$ transition from the ground electronic $N$ state to the valence excited $V$ state [11]. As a prototypical system for ultrafast internal conversion, the VUV excitation and subsequent dissociation of ethylene have been the subject of extensive studies ([12,13] and work cited therein). Time-resolved measurements to date ([4,14] and references therein) have been limited to studies employing multiphoton ionization that observe only the decay of the excited population and not the formation of the products. The VUV/XUV pump–probe apparatus allowed us, for the first time, to directly time resolve the formation of the $H_2$ dissociation product. The $H_2^+$ signal from $C_2H_4$ is shown in Fig. 3(b) along with the decay of the $C_2H_4^+$ signal due to depletion of the ground state population. The $C_2H_4^+$ signal rapidly decays as the nuclear wave function moves away from the Franck–Condon region, and its overlap with bound states of the cation becomes negligible. On longer time scales, $H_2$ elimination occurs in the vibrationally hot electronic ground state [12,13]. We fit the $H_2^+$ signal to an exponential rise convolved with the FIR and obtain a time constant of $184 \pm 5$ fs. In addition to the molecular elimination, more details of the ultrafast dynamics can be extracted from the full ion fragment mass spectrum [15].

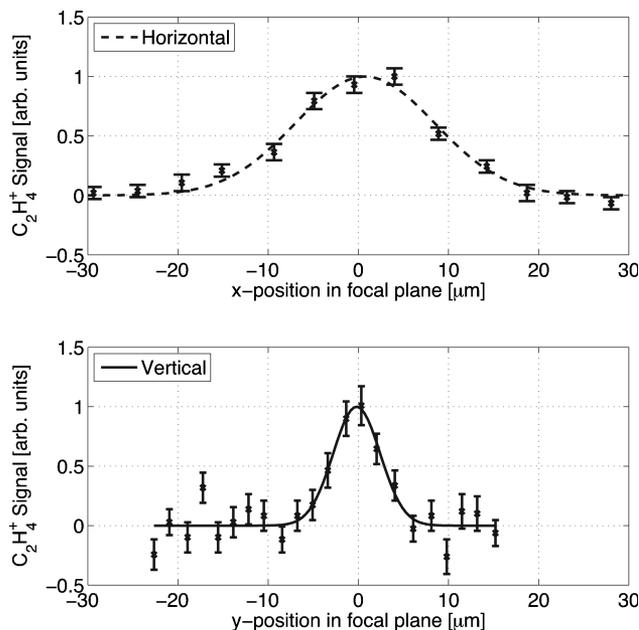

Fig. 2. Split-mirror foci spatial autocorrelation: multiphoton ionization of $C_2H_4$ as a function of the overlap of the two SMI foci in the horizontal (top) and vertical (bottom) directions. Gaussian fits give a fifth-harmonic focus size of $4~\mu m \times 12~\mu m$.

We expect the techniques described here to find broad applications in femtosecond spectroscopy, just as static measurements with VUV and XUV light sources have been indispensible for our understanding of molecular physics. The apparatus has proven to be a versatile tool. In addition to the results presented here, we have successfully conducted experiments in a fifth-harmonic pump/fifth-harmonic probe configuration with $25 \pm 7$ fs resolution by using the 160 nm spectral bandpass filter in both arms of the SMI and generating harmonics with 8 Torr of Ar in the gas cell. A fifth-harmonic pump/nineteenth-harmonic probe configuration using 3.2 Torr Kr in the gas cell and a 0.16-µm-thick Al filter/Mg:SiC multilayer mirror in the probe arm demonstrated $13^{+9}_{-0}$ fs resolution [5]. We comment that our choice of pump and probe was not limited by the flux from the HHG source. We estimate the ability to ionize up to 3% of the molecules in the focal volume with the XUV probe, indicating that the available pump wavelengths are not restricted to the intense fifth harmonic. As higher repetition rate terawatt-class lasers become available, more detailed information can be obtained with coincidence or charged particle imaging techniques not yet feasible with our 10 Hz system.

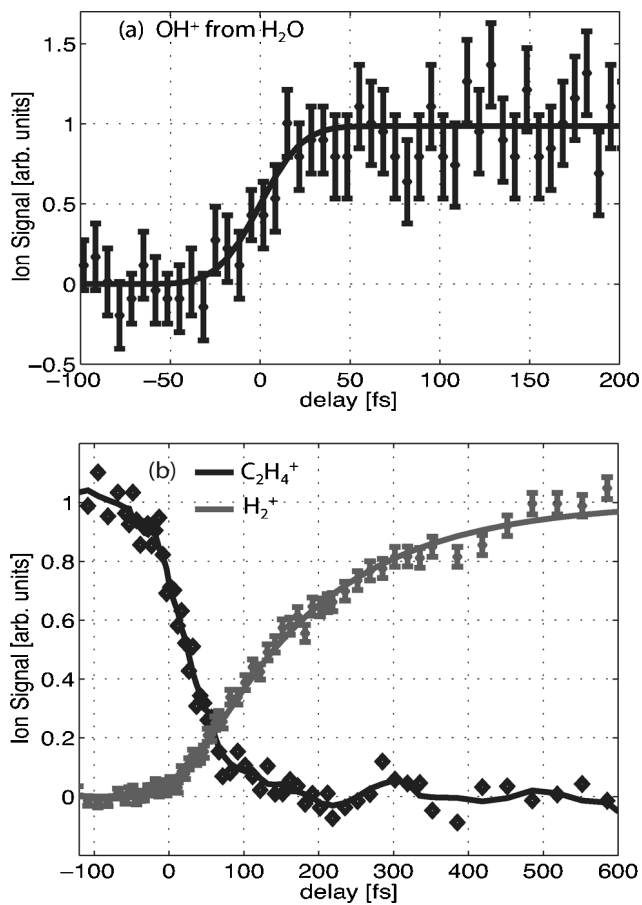

Fig. 3. (Color online) Time-resolved photodissociation: (a) time-resolved $OH^+$ yield for $H_2O$ target gas. The $OH^+$ yield is fit with an error function fit (solid curve) to give an upper limit for the FIR of $44^{+15}_{-10}$ fs FWHM. (b) Red points, $H_2^+$ yield from $C_2H_4$; red curve, exponential rise with 184 fs time constant; black points, $C_2H_4^+$ signal; black curve, adjacent point smoothed. Note the different $x$ scales for (a) and (b).

This work was supported by the United States Department of Energy (DOE) Office of Basic Energy Sciences, under contract numbers DE-AC02-05CH1123 and DE-FG-52-06NA26212. A. M. Stooke gratefully acknowledges the full support of the Fannie and John Hertz Foundation. We thank A. Stolow, H. Tao, and T. J. Martinez for helpful discussions. We acknowledge C. Caleman, M. Bergh, H. Merdji, and M. P. Hertlein for help with the apparatus in its early stages.